\documentclass[aps,prl,showpacs,twocolumn,preprintnumbers,amsmath,amssymb]{revtex4-1}

\usepackage{graphicx}
\usepackage{dcolumn}
\usepackage{bm}
\usepackage{color}
\usepackage{soul}

\def\st#1{\unskip\relax}

\newcommand{\sech}[1]{\textrm{sech}\left(  #1\right)}

\begin{document}

\title{The multi-resonant Lugiato-Lefever model}

\author{Matteo Conforti$^{1}$ and Fabio Biancalana$^2$}
\affiliation{
$^1$Univ. Lille, CNRS, UMR 8523 - PhLAM - Physique des Lasers Atomes et Mol{\'e}cules, F-59000 Lille, France\\
$^2$School of Engineering and Physical Sciences, Heriot-Watt University, EH14 4AS Edinburgh, UK}


\date{\today} 
 
\begin{abstract} 
We introduce a new model describing multiple resonances in Kerr optical cavities. It perfectly agrees quantitatively with the Ikeda map and predicts complex phenomena such as super cavity solitons and coexistence of multiple nonlinear states.
\end{abstract}

\maketitle

Optical resonators featuring Kerr media display a wealth of phenomena, encompassing frequency combs \cite{Kippenberg2011}, cavity solitons \cite{Herr2013,Leo2010} and instabilities \cite{Copie2016}, which are being intensively studied in view of their high impact applications \cite{Kippenberg2011}. Given the complexity and the diversity of the physical phenomena, deriving  simple, accurate and efficient models is of paramount importance. The workhorse for the description of nonlinear cavity dynamics is the celebrated Lugiato-Lefever equation (LLE) \cite{Lugiato1987,Haelterman1992}, which allows for deep theoretical insight and fast and accurate numerical modelling. 
Despite the fact that the LLE holds valid well beyond the mean-field approximation under which it has been historically derived, it is not capable of modelling all the phenomena of interest. Indeed, LLE can model the evolution of only one Fabry-P\'erot mode, that corresponds to a {\em single resonance}. Phenomena not captured by LLE range from "super cavity solitons" (SCSs) \cite{Hansson2015} to the coexistence of stable modulational instability (MI) patterns and solitons, observed in Ref. \cite{Anderson2017}. The complete and exact dynamical scenario can be reproduced by the famous Ikeda map \cite{Ikeda1980}, but this model does not give any reasonable physical insight owing to its complex mathematical structure. A model sharing the accuracy of the Ikeda map and the simplicity of LLE will be of paramount importance in the design of the next-generation of resonators \cite{Kartashov2017}. A reliable model capable to reproduce quantitatively the results of the Ikeda map and the latest experiments \cite{Anderson2017} is still missing. 
In this Memorandum, we derive rigorously a multi-resonant LLE system, which agrees \emph{quantitatively} with the Ikeda map. Each Fabry-P\'erot resonance is described by an LLE-type equation, which is coupled to the others in a nontrivial way. An arbitrary number of resonances can be treated, making the model highly flexible and scalable to any situation of experimental interest.
For definiteness, we consider a fiber ring cavity, but the method can be of course straightforwardly applied to micro-resonators.

We start from the Ikeda map in dimensional units: 
\begin{align}
\label{bc} E^{(n+1)}(Z=0,T)=\theta E_{in}+\rho e^{i\phi_0}E^{(n)}(Z=L,T),\\
\label{nls} i\frac{\partial E^{(n)}}{\partial Z}-\frac{\beta_2}{2}\frac{\partial^2 E^{(n)}}{\partial T^2}+\gamma|E^{(n)}|^2E^{(n)}=0,\;\; 0<Z<L.
\end{align}
$E^{(n)}$ is the electric field envelope at the $n$-th round-trip (measured in $\sqrt{W}$), $P_{in}=|E_{in}|^2$ is the input pump power, $\rho^2$, $\theta^2$ are respectively  the power reflection and transmission coefficients of the coupler, and $\phi_0=\beta_0L$ is the linear cavity round-trip phase shift. For simplicity we lump all the losses in the boundary condition, with $1-\rho^2$ describing the total power lost per round-trip. $Z$ measures the propagation distance inside the fiber of length $L$, and $T$ is time in a reference frame traveling at the group velocity of the pulse.
%
To proceed, we note that the map above can be replaced -- without loss of generality -- with a single equation where the boundary conditions are explicitly incorporated in an NLSE-type equation  of an ``unfolded cavity''. Specifically, using the Dirac delta comb to model the periodic application of the boundary conditions, and using the identity $\sum_n\delta(Z-nL)=\frac{1}{L}\sum_n e^{inkZ}$ (where $k=2\pi/L$) and letting $Z \in[0,+\infty)$ (``unfolded" cavity), we arrive at the equation:
\begin{align}\label{main}
\nonumber i\frac{\partial E}{\partial Z}-\frac{\beta_2}{2}\frac{\partial^2 E}{\partial T^2}+\gamma|E|^2E&=\\ \frac{i\theta}{L}E_{in}\sum_n&e^{i(nk-\beta_0)Z} +i\frac{\rho -1}{L}E\sum_ne^{inkZ}.
\end{align}
Equation (\ref{main}) is the NLSE forced by two combs with equal wave-number spacing $k$ and a relative  shift $\beta_0$. A conceptually different model was derived very recently following a single-equation approach \cite{Kartashov2017}, but crucially it is not based on the exact Ikeda map and it is still far too complicated to allow a deep physical understanding (e.g. stationary solutions like CSs cannot be found even numerically). 
The solution of \eqref{main} can be written as a sum of slowly-varying envelopes, which modulate the longitudinal Fabry-P\'erot modes of the cavity. We assume that $N=N_R+N_L+1$ Fabry-P\'erot resonances are efficiently excited: $E(Z,T)=\sum_{n=-N_R}^{N_L}E_n(Z,T)e^{iknZ}$, where $N_L$ (or $N_R$), are the number of modes corresponding to a resonance to the left towards smaller detuning (or to the right towards bigger detuning), of the central resonance denoted $n=0$.
By collecting exponentials oscillating with the same wave-number, we arrive at the following compact and general expression, consisting of $N$ coupled LLEs (CLLEs):
\begin{align}\label{CLLE}
\nonumber i\frac{\partial U_n }{\partial Z}-&\frac{\delta_n}{L}U_n-\frac{\beta_2}{2}\frac{\partial^2 U_n}{\partial T^2}+\gamma\sum_{p=-N_R}^{N_L}\sum_{q=q_{min}}^{q_{max}}U_pU_qU^*_{p-n+q}=\\ &=i\frac{\theta}{L} E_{in} -i\frac{\alpha}{L} \sum_{p=-N_R}^{N_L}U_p ,\;\;\;(n=-N_R,\ldots,N_L)
\end{align}
where $\alpha=1-\rho$, $U_n=E_n\exp[i\delta_0Z/L]$, 
 $q_{min}=\max\{-N_R,n-p-N_R\}$, $q_{max}=\min\{N_L,n-p+N_L\}$, $\delta_n=\delta_0+2\pi n$. The conditions on the integers $q_{min,max}$ select only the correct nonlinear couplings.
%
%
The cavity detuning from the central mode is defined as $\delta_0=mk-\beta_0$, $m=\arg\min_n|nk-\beta_0|$, entailing $-\pi\le \delta_0\le\pi$, which is consistent with the $2\pi$ periodicity of the Ikeda map. 
%
If we assume that only one mode is excited in the cavity ($N_R=N_L=0$) we recognize in \eqref{CLLE} the standard, single-resonance LLE. It is worth noting that our CLLE (\ref{CLLE}) {\em do not use the mean-field approximation}. In the standard single-resonance LLE this approximation must be used since the field cannot change much over one roundtrip. However in our formulation the mean-field approach is not used, since the field is allowed to change arbitrarily fast and oscillate strongly, due to the presence of a large number of resonances.
%

\begin{figure}[htbp]
\centering
\fbox{\includegraphics[width=0.95\linewidth]{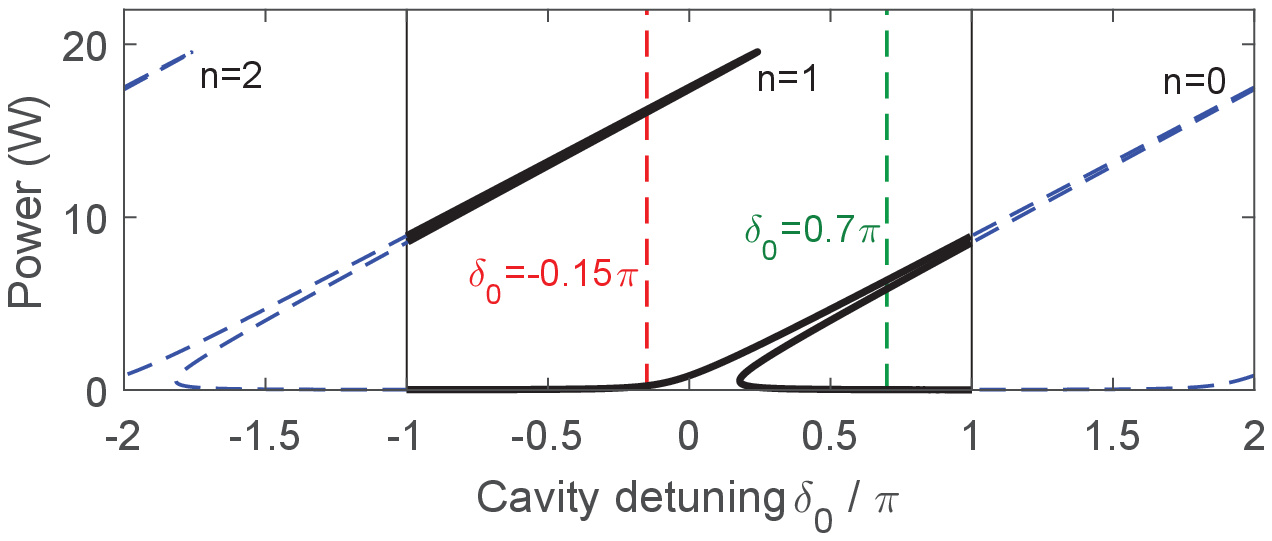}}
\caption{Steady-state response of the cavity from Ikeda map (dashed blue) and CLLEs Eq. (\ref{CLLE}) (solid black) with $N=3$ resonances ($N_L=2,N_R=0$). Parameters $L=300$ m, $\alpha=0.0619$, $\theta^2=0.05$, $\beta_2=-22$ ps$^2$/km, $\gamma=1.2$ W$^{-1}$km$^{-1}$, $P_{in}=|E_{in}|^2=1.5$ W.}
\label{fig1}
\end{figure}

In order to test our model, we simulate the fiber ring resonator described in Ref. \cite{Anderson2017} (parameter listed in the caption of Fig. \ref{fig1}).
We found that $N=3$ resonances ($N_L=2,N_R=0$) describe perfectly the full range of detuning $-\pi\le\delta_0\le\pi$ for $P_{in}=1.5$ W.
Figure \ref{fig1} shows the tilted cavity resonances obtained by solving the stationary ($\partial/\partial Z=\partial/\partial T=0$) CLLEs Eq. (\ref{CLLE}) via a Newton-Raphson method (solid black curve). The agreement with the Ikeda map (dashed blue curve) is {\em perfect} over the full range $-\pi\le\delta_0\le\pi$.
\begin{figure}[htbp]
\centering
\fbox{\includegraphics[width=0.95\linewidth]{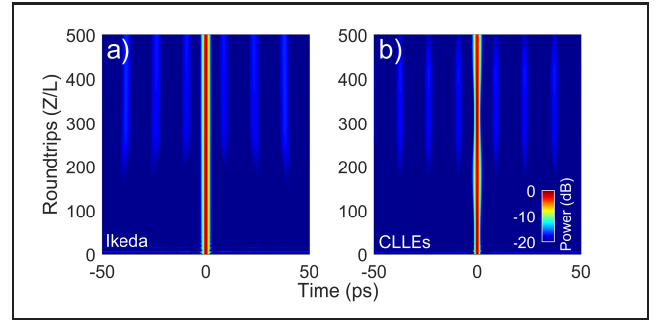}}
\caption{Coexistence between CS and MI pattern at $\delta_0=-0.15\pi$ (vertical dashed red line in Fig. \ref{fig1}). Input for Ikeda  map $|E^{(0)}(Z=0)|^2=P_S\sech{T/T_S}^{2}$; for CLLEs $|U_1(Z=0)|^2=P_S\sech{T/T_S}^{2}$, $U_{0,2}(Z=0)=0$. Small white noise is added to the initial condition. Parameters: $P_S=2(\delta_0+2\pi )/(\gamma L)$, $T_S^2=|\beta_2|L/(2(\delta_0+2\pi ))$ \cite{Hansson2015}}
\label{fig2}
\end{figure}
At a detuning $\delta_0=-0.15\pi$  (dashed red vertical line in Fig. \ref{fig1}), we expect the coexistence of a stable MI pattern, given by the resonance $n=0$, and a SCS supported by the resonance $n=1$. This is confirmed by numerical solution of the Ikeda map in Fig. \ref{fig2}, which matches remarkably well with the CLLEs. 
At a detuning $\delta_0=0.7\pi$  (dashed green vertical line in Fig. \ref{fig1}), the SCS associated to $n=1$ resonance still exists, with a higher power proportional to $\delta_0+2\pi $. Moreover a conventional CS of lower power proportional to $\delta_0$ is supported by the fundamental $n=0$ cavity resonance. This coexistence is confirmed by numerical solution of the Ikeda map reported in Fig. \ref{fig3}(a), corresponding to an initial field composed of a simple sum of the approximated CS and SCS separated by 20 ps in time. A similar picture (not shown) is obtained from CLLEs. A Newton-Raphson solution of steady CLLEs shows that the CS has its power concentrated in the component $n=0$ [Fig. \ref{fig3}(b)], whereas the SCS resides predominantly in the component $n=1$ [Fig. \ref{fig3}(c)].

\begin{figure}[htbp]
\centering
\fbox{\includegraphics[width=0.95\linewidth]{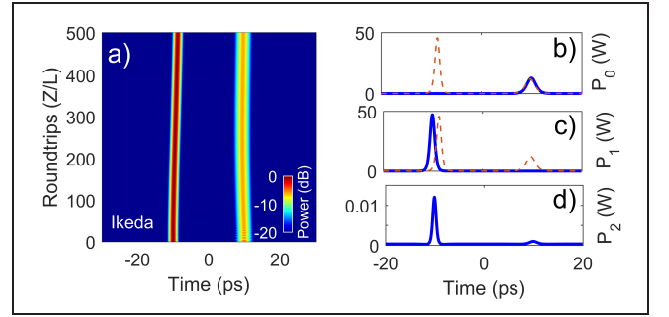}}
\caption{Coexistence between CS and SCS  at $\delta_0=0.7\pi$ (dashed red line in Fig. \ref{fig1}). (a): Intracavity power from Ikeda  map with approximated input CS+SCS \cite{Hansson2015} separated by 20 ps. (b,c,d): Power profiles ($P_n=|U_n|^2$) from solution of steady CLLEs (blue solid lines). Red dashed line is the output from Ikeda map.}
\label{fig3}
\end{figure}

To conclude, we have derived a model based on coupled LLEs which allows for the accurate description of Kerr optical cavities when several Fabry-P\'erot resonances interact. It is extremely accurate even when including a small number of modes, and will allow to acquire great physical insight thanks to its simplicity.  

\section*{Acknowledgments}
The authors gratefully acknowledge fruitful discussions with M. Erkintalo, S. Coen and S. Murdoch.

\end{document}